\documentclass[11pt,twoside]{article}

\usepackage{asp2004}
\usepackage{epsf}
\usepackage{psfig}
\usepackage{lscape}

\begin{document}
\title{Mid-IR selected Quasars in the First Look Survey}   
\author{A.Petric$^{1,2}$, M. Lacy$^1$, L.J. Storrie-Lombardi$^1$, A. Sajina$^1$, L. Armus$^{1}$, G. Canalizo$^{3}$, S. Ridgway$^4$}   
\affil{$^1$Spitzer Science Center, California Institute of Technology, Mail Code 220-6, 1200 East Clifornia, Blvd., Pasadena, CA 91125, $^2$Columbia University, $^3$UC Riverside,$^4$John Hopkins U.}    

\begin{abstract} 
We present a preliminary investigation of the spectral energy distributions (SEDs), and star-formation properties of a sample of Mid-IR selected Quasars. The mid-infrared SEDs of our objects are consistent with that expected
from clumpy torus models. At longer infrared wavelengths, the radio to
infrared ratios of several objects are consistent with those of star-forming galaxies.
\end{abstract}
\vskip -5in
\section{Introduction}
	A central issue in the study of the formation and evolution of
galaxies is the connection between the central back hole and the
surrounding bulge stars. Observations of the dynamics of stars and gas
in the nuclear regions of nearby galaxies suggest that the
overwhelming majority of spheroidal galaxies in the local Universe
contain massive black holes (BH), and that the mass of the central black
hole correlates with the velocity dispersion of the stars in the
spheroid (e.g. Gebhardt et al.~2000). This suggests a fundamental relation between the formation of massive BH and the
stellar content of galaxies. Valid tests of the coeval growth theories are contingent upon a deeper understanding
 of the processes associated with the the formation of the spheroid and those close to the
 galactic nucleus which regulate the growth of the black hole by accretion. Such understanding can be partially gained 
through the study of large numbers of quasars, especially the obscured population, some members of which may represent
quasars obscured by star-forming host galaxies.    
\vskip -0.3in
\section{Sample Selection}
Our sample contains 77  MID-IR selected AGN
(Lacy et al.~2004,2005) for which we have IRAC, and 
MIPS 24/70 $\mu$m data. VLA 1.4GHz and ugrizJHK data are available for the majority of the
sample. The sources were selected from the 
Spitzer First Look Survey (e.g. Lacy et al.~2004) and  from the Spitzer Wide-area InfraRed
 Extragalactic (SWIRE) Legacy Survey (Lonsdale et al.~2003).
The sources presented here were chosen on the basis of both their IRAC colors
and their MIPS 24 $\mu$m fluxes.  The IRAC color selection, discussed e.g. in Lacy et al. (2004,2005), Stern (2005) 
 has been derived empirically, is supported by modeling of ISO spectra by Sajina et al. (2005), and adds
greatly to the completeness and reliability of optical AGN selections (e.g. Lacy et al. 2006, Richards et al. 2006, Siana et al. 2006). To select the objects presented
here we also required that they should have 24 $\mu$m fluxes  greater than $\sim$ 4mJy. Selection based on a 24$\mu$m rather than the 8$\mu$m-based
 selection of Lacy et al. (2004) is a more sensitive to highly-reddened AGN, and QSO's at high redshift  where 
the strong k-correction on the mid-infrared dust emission means that they drop out of flux-limited samples selected at shorter wavelengths. The selected sources were followed up spectroscopically, these observations are thoroughly discussed in Lacy et al. (2006). Figure 1 shows the main classes of quasars found in the MIR selected sample. 
\begin{figure}[!ht]
\plottwo{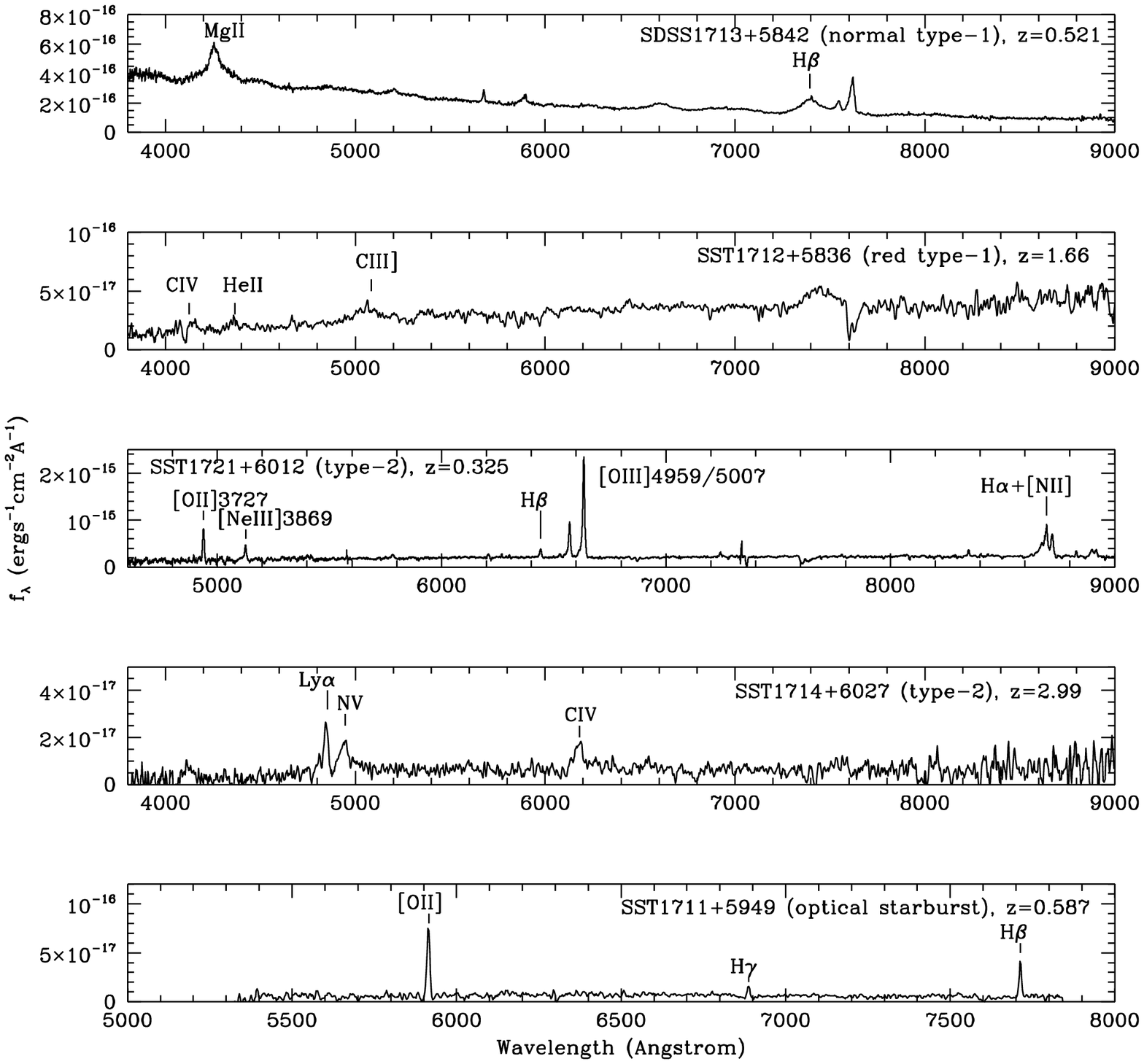}{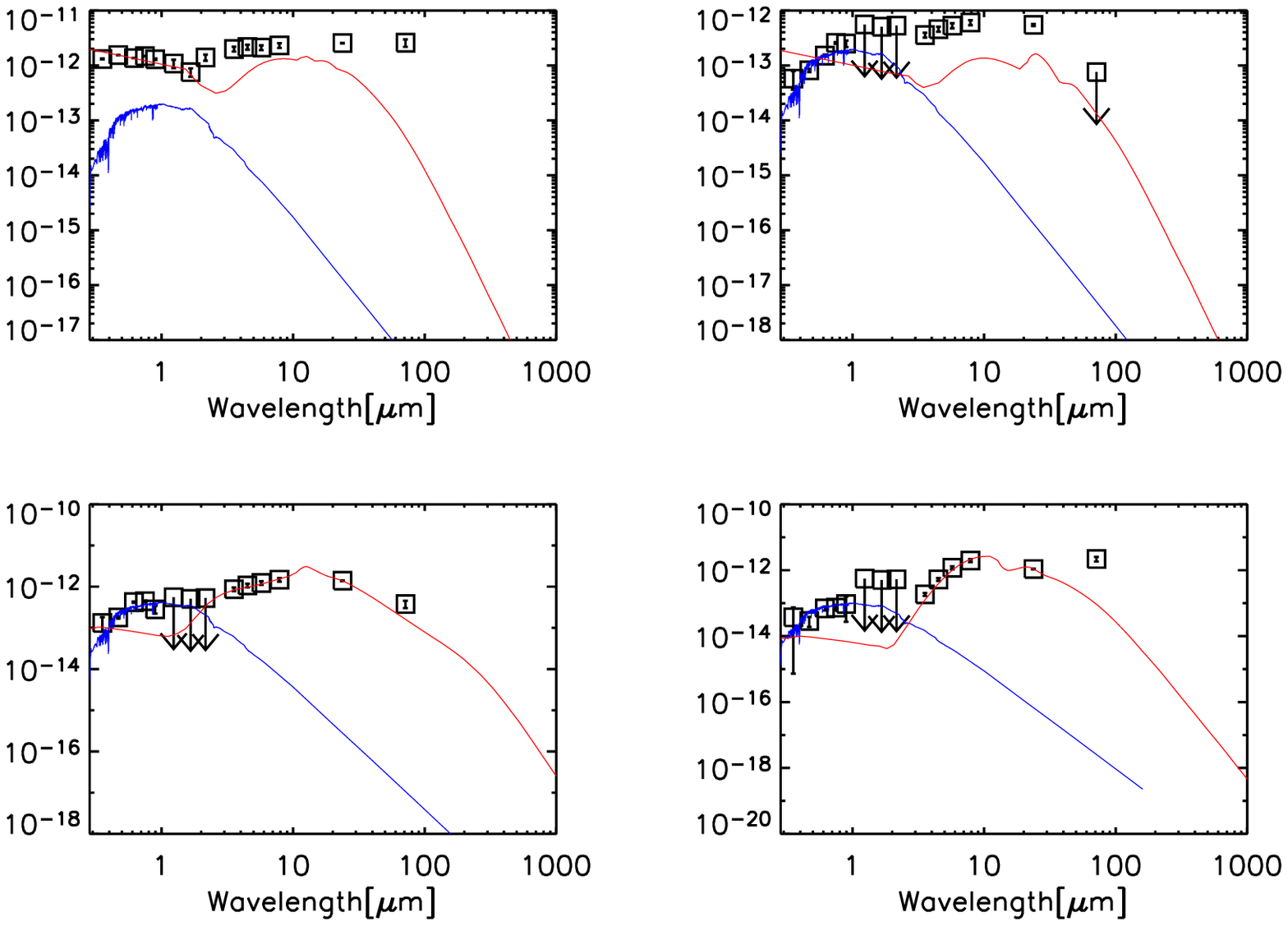}
\caption{On the left we show five optical spectra representative of the types of AGN's detected with the MIR selection. The first panel in this figure is a typical
type 1 QSO with strong permitted lines and FeII emission in the 4000-5000\AA range. Panel 2 shows a reddened quasar with broad
optical lines. The third and fourth plots are type 2 quasars. These are objects with high-ionization narrow emission lines
and it is only recently that these type of quasars have been admitted as the bigger and brighter counterpart of Seyfert IIs
(e.g. Zakamska 2003,2006). The last panel shows a type 3 source,( Leipski et al. 2005). These latter sources
lack any obvious signs of AGN activity in their optical spectra.
On the right we show SED's and best model fits for the spectra shown in the top figures, panels 1,2,3,and 5 respectively. The $X$-axis gives the observed wavelength
while the $Y$-axis is in arbitrary units but is proportional to $\nu ~{\rm{F}}_{\nu}$. The blue curve 
shows the estimated host contribution from a 5~Gyr stellar population. The red curve shows the AGN-torus emission as indicated
from our fits of the Nenkova et al. (2002) models. }
\end{figure}
\section{Estimated Sample Properties}
{\bf{Modeling the SEDs}}
 ~We fit models of a clumpy torus developed by Nenkova et al. (2002) and our preliminary result is that they fit well our spectra, (e.g. their estimated inclinations seem to match these expected from the optical spectroscopy). 
A host galaxy contribution seems to be needed for all but a few type 1 sources.

\noindent{\bf{Star formation}}
~One of the mechanisms invoked to maintain a stable torus is  nuclear star formation (e.g.Wada \& Norman 2002). Also, 
the theory of coeval and supermassive black hole growth implies that objects with accreting AGN 
are undergoing star-formation in the host galaxy and/or the torus. One method to obtain an indication whether
massive star formation occurs in a source is to check if the IR to radio flux ratios match those of nearby non-AGN
star-forming galaxies. In the local universe, star-forming sources from optically, IR or radio selected samples follow a very tight linear 
relation between the radio-continuum and FIR luminosities, with only a factor of two scatter around linearity over
four orders of magnitude in luminosity (Condon 1992; Yun, Reddy, \& Condon 2001) The radio to FIR correlation has been quantified through the $q$
parameter, defined as:
$$q~= \rm ~log\left({{L_{FIR}}\over{3.75~\times~10^{12}~W \rm{m}^{-2}}}\right)~-~log\left({F_{1.4}\over{Wm^{-2} Hz^{-1}}}\right)$$ where L$_{\rm{FIR}}$
corresponds to the FIR luminosity in solar units between 40 and 120 $\mu$m and F$_{1.4}$ is
the radio flux at 1.4~GHz in W~m$^{-2}$~Hz$^{-1}$.  In a study of 2000 IRAS-selected galaxies,  Yun, Reddy \& Condon
(2001) find a mean $q$ value of 2.34, and that the
range ($q = 1.64$ to 3) corresponds to star forming galaxies, while
significantly lower $q$ values imply that the dominant emission is from the AGN.

We estimate the restframe 60 $\mu$m luminosity using the measured, or upper-limit spectral index between 70 and 24 $\mu$m. Frayer et al. (2006) find a ratio of 2.3 between the 100 and 60 $\mu$m flux
of 2.3. Condon (1992) relates the 60 (S$_{60})$ and 100(S$_{100}$) $\mu$m fluxes in $Jy$s to the total FIR emission in {W m$^{2}$}: $S_{\rm{FIR}}~=~1.26e-4(2.58{\rm{S}}_{60}~+~{\rm{S}}_{100})$.We find that several of our AGNs tend to follow the FIR-radio correlation for nearby star-forming sources, as
previously determined by other workers e.g. Sopp \& Alexander (2001). 

The 70 to 24 $\mu$m flux ratios provide another way to estimate the contribution of either host or 
starformation in and outside the torus to the observed emission since these values are expected
to range between 0.2 and 2 for AGN sources, and an excess of 100  $\mu$m emission is seen in 
star forming objects (e.g. Canalizo \& Stockton 2001). Figure 2b showing the spectral index at 8-24 $\mu$m plotted against that from 24-70 also suggests that the type 3 (open circles) are more likely to have an IR-excess, due perhaps to more cold dust, and star-formation. 
\begin{figure}[!ht]
\plottwo{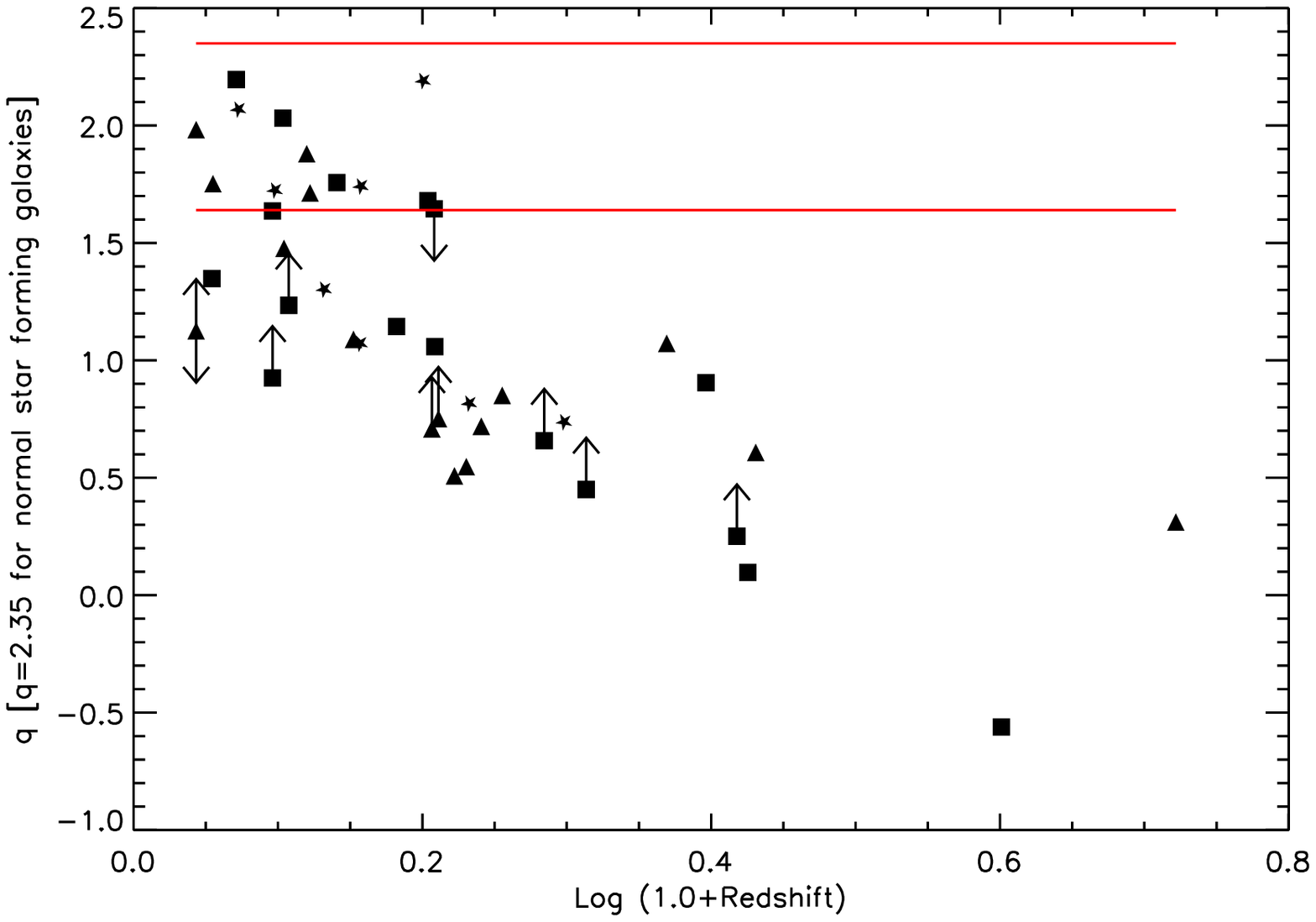}{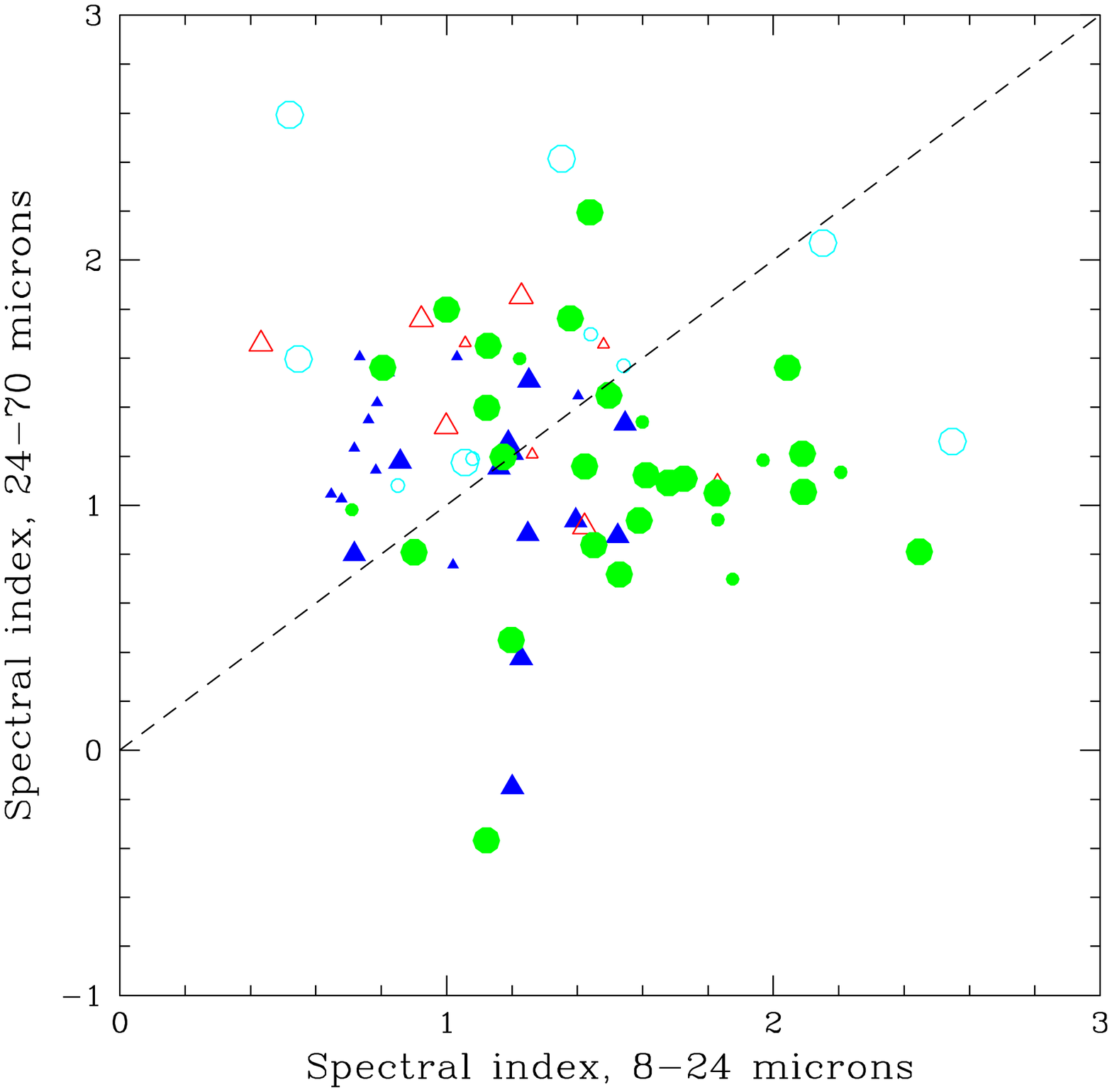}
\caption{The estimated FIR and radio properties of MIR selected AGN with triangles, squares, and stars
 representing type 1, 2 and 3 quasars respectively. The plot on the left shows the $q$ parameter versus redshift for our sample. The average $q$ value
for nearby star-forming galaxies is 2.35. The figure on the right shows the spectral index at 8-24 $\mu$m plotted against that from 24-70,
with the IR-excess objects plotted above the line. Closed triangles are type-1 objects, open triangles type-2s, closed circles reddened 
type 1s and open circles type 3s. The smaller symbols show upper limits at 70 $\mu$m.} 
\end{figure}
\vskip -0.2in
\section{ Conclusions and Future Work}
Initial SED modeling suggests that the emission from these objects are consitent with what is expected from clumpy
torus models (Nenkova et al. 2002) with contributing light from a host
galaxy with an old stellar population ($\sim$5 Gyr).
We are currently investigating whether adding a younger population of stars changes our results.
We estimate  that several of our obscured sources have similar FIR to radio properties as those of star-forming nearby galaxies.
 Scheduled HST observations will allow us to image the host galaxies of several
type 2 and type 3 sources and relate them to their infrared spectra, and to observations of optically selected type-2 QSOs(e.g. Zakamska et al. 2006).
 Available and proposed IRS spectroscopy, together with more sophisticated model fitting, will give us
further insight into the star-formation processes, origin of extinction, and physical properties
of the dust associated with the AGN.

\end{document}